\documentclass[conference]{IEEEtran}
\IEEEoverridecommandlockouts
\usepackage{cite}
\usepackage{amsmath,amssymb,amsfonts}
\usepackage{algorithmic}
\usepackage{graphicx}
\usepackage{textcomp}
\usepackage{xcolor}

\usepackage{cleveref}
\usepackage{subfig}
\usepackage{gensymb}

\usepackage{balance}

\begin{document}

\title{On the Sustainability of Lightweight Cryptography Based on PUFs Implemented on NAND Flash Memories Using Programming Disturbances
\thanks{This work has been funded by the German Research Foundation -- Deutsche Forschungsgemeinschaft (DFG), as part of the Projects ``PUFMem: Intrinsic Physical Unclonable Functions from Emerging Non-Volatile Memories'' (project number 440182124) and ``NANOSEC: Tamper-Evident PUFs based on Nanostructures for Secure and Robust Hardware Security Primitives'' (project number 439892735) of the Priority Program ``Nano Security: From Nano-Electronics to Secure Systems'' (SPP 2253).}
}

\author{\IEEEauthorblockN{Nikolaos Athanasios Anagnostopoulos\IEEEauthorrefmark{1}\IEEEauthorrefmark{2}, Yufan Fan\IEEEauthorrefmark{3}, Muhammad Umair Saleem\IEEEauthorrefmark{4},\\Nico Mexis\IEEEauthorrefmark{1}, Florian Frank\IEEEauthorrefmark{1}, Tolga Arul\IEEEauthorrefmark{1}\IEEEauthorrefmark{2}, Stefan Katzenbeisser\IEEEauthorrefmark{1}}

\IEEEauthorblockA{\IEEEauthorrefmark{1}University of Passau, Faculty of Computer Science and Mathematics, Innstraße 43, 94032 Passau, Germany\\
Emails: \{Nikolaos.Anagnostopoulos, Florian.Frank, Tolga.Arul, Stefan.Katzenbeisser\}@uni-passau.de}
\IEEEauthorblockA{\IEEEauthorrefmark{2}Technical University of Darmstadt, Computer Science Department, Hochschulstraße 10, 64289 Darmstadt, Germany\\
Emails: \{na45tisu, arul\}@rbg.informatik.tu-darmstadt.de}
\IEEEauthorblockA{\IEEEauthorrefmark{3}Technical University of Darmstadt, Department of Electrical Engineering and Information Technology,\\Merckstraße 25, 64283 Darmstadt, Germany\\
Email: yufan.fan@nt.tu-darmstadt.de}
\IEEEauthorblockA{\IEEEauthorrefmark{4}Technical University of Darmstadt, Computer Science Department, Hochschulstraße 10, 64289 Darmstadt, Germany\\
Email: umsach-contact@protonmail.com}
}

\maketitle

\begin{abstract}
In this work, we examine the potential of Physical Unclonable Functions (PUFs) that have been implemented on NAND Flash memories using programming disturbances to act as sustainable primitives for the purposes of lightweight cryptography. In particular, we investigate the ability of such PUFs to tolerate temperature and voltage variations, and examine the current shortcomings of existing NAND-Flash-memory PUFs that are based on programming disturbances as well as how these could potentially be addressed in order to provide more robust and more sustainable security solutions.
\end{abstract}


\section*{Brief Overview of this Work}

Flash-memory-based Physical Unclonable Functions (PUFs) have recently been proposed in the relevant literature~\cite{jiaxia,8993414,prabhu2011extracting,6234403,7939086} as a lightweight and sustainable security primitive, because their implementation and operation are rather cost-efficient. More specifically, only lightweight software is required for their operation, and they either may not require any hardware addition, as Flash memories are often inherent parts of computing systems, e.g., Internet-of-Things (IoT) devices, or may be reusable in different systems, as they are also often found in removable modules. 

In general, PUFs are physical objects, such as hardware, which utilise minor manufacturing variations, in order to provide a rather unique output for a specific input under partcular conditions. In the case of NAND-Flash-memory-based PUF that utilises programming disturbances, certain pages of the Flash memory are programmed rapidly and repeatedly causing a unique error pattern to appear in nearby pages.

\begin{figure}[htb!]
  \vspace{-5pt}
  \centering
  \includegraphics[width=0.55\linewidth]{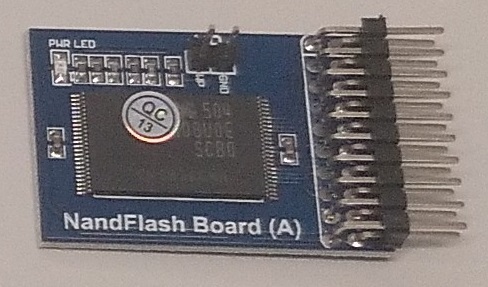}
  \caption{A photo of the Waveshare NandFlash Board (A).}
  \label{fig:NANDFlashboard}
  \vspace{-5pt}
\end{figure}

In this work, we will examine the quality of the responses of such Flash-memory-based {PUF} that have been implemented on multiple instances of the Waveshare NandFlash Board (A), which is a removable external Flash memory module that incorporates a 1-Gbit Samsung K9F1G08U0E NAND Flash memory, as shown in~\Cref{fig:NANDFlashboard}. Following the example of previous works regarding NAND-Flash-memory-based PUFs that utilise programming disturbances~\cite{prabhu2011extracting,jiaxia,8993414}, we also assume that each individual page of the relevant NAND Flash memory that is affected by the rapidly repeated programming of its nearby pages constitutes a different instance of this PUF. Therefore, in order to examine the sustainability of this PUF type, we examine the quality characteristics of the responses of each such PUF, i.e., the values of the cells of each Flash memory page used as a PUF instance, after the operation of this PUF under adverse environmental conditions, such as temperature and voltage variations.

Each Waveshare NandFlash Board~(A) used in our work was controlled using an ST Microelectronics STM32F429I Discovery (STM32F429I-DISC1) board, to which this Flash board had been connected using a Waveshare Open429Z-D Standard, an STM32F4 expansion/development board for the ST Microelectronics STM32F429I Discovery (STM32F429I-DISC1) board. A single Flash memory block was fully erased and, thus, the initial bit pattern of cells utilised in the relevant {PUF} response was \texttt{0xFF} (all ones), while the other cells of the same block were rapidly and repeatedly programmed with the bit pattern \texttt{0x00} (all zeros), for 10,000 programming cycles or until at least 2040 addresses (corresponding to 1 {B} each) have had at least one bit flip each, whichever condition was fulfilled first. Thus, in case the latter condition holds true, in each of the 2040 bytes of the relevant {PUF} page, which consists of 2048 bytes in total, there will be at least one bit flip, i.e., all but one of the page's bytes will contain at least one bit flip.

\begin{figure}[tb!]
\centering
\includegraphics[width=0.9\linewidth]{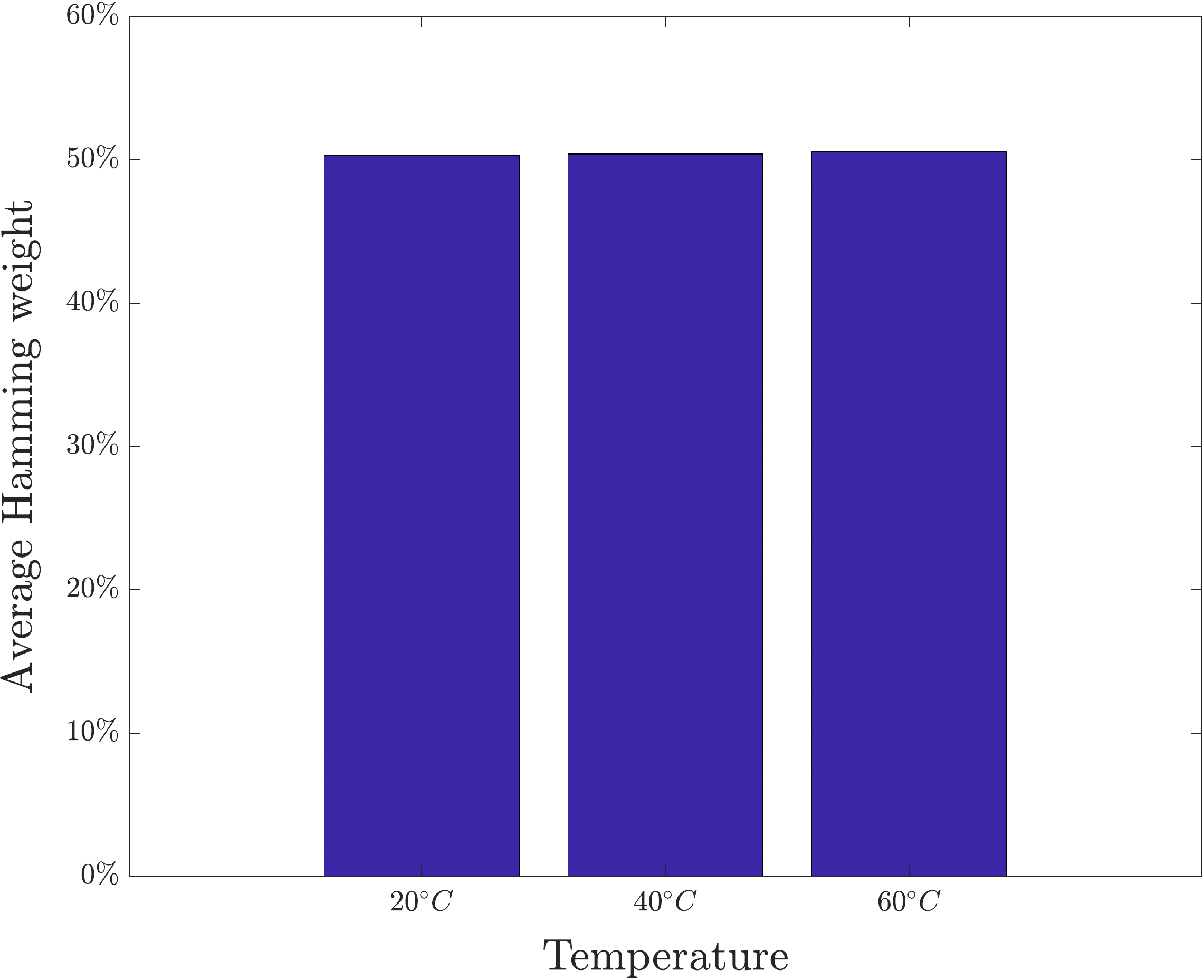}
\vspace{-2.5pt}
\caption{Overall average fractional Hamming weight value of the examined NAND-Flash-memory-based {PUF}, for each value of the ambient temperature tested.}
\label{fig:flashtempHWoverall}
\vspace{-5pt}
\end{figure}

\begin{figure}[tb!]
\centering
\includegraphics[width=0.9\linewidth]{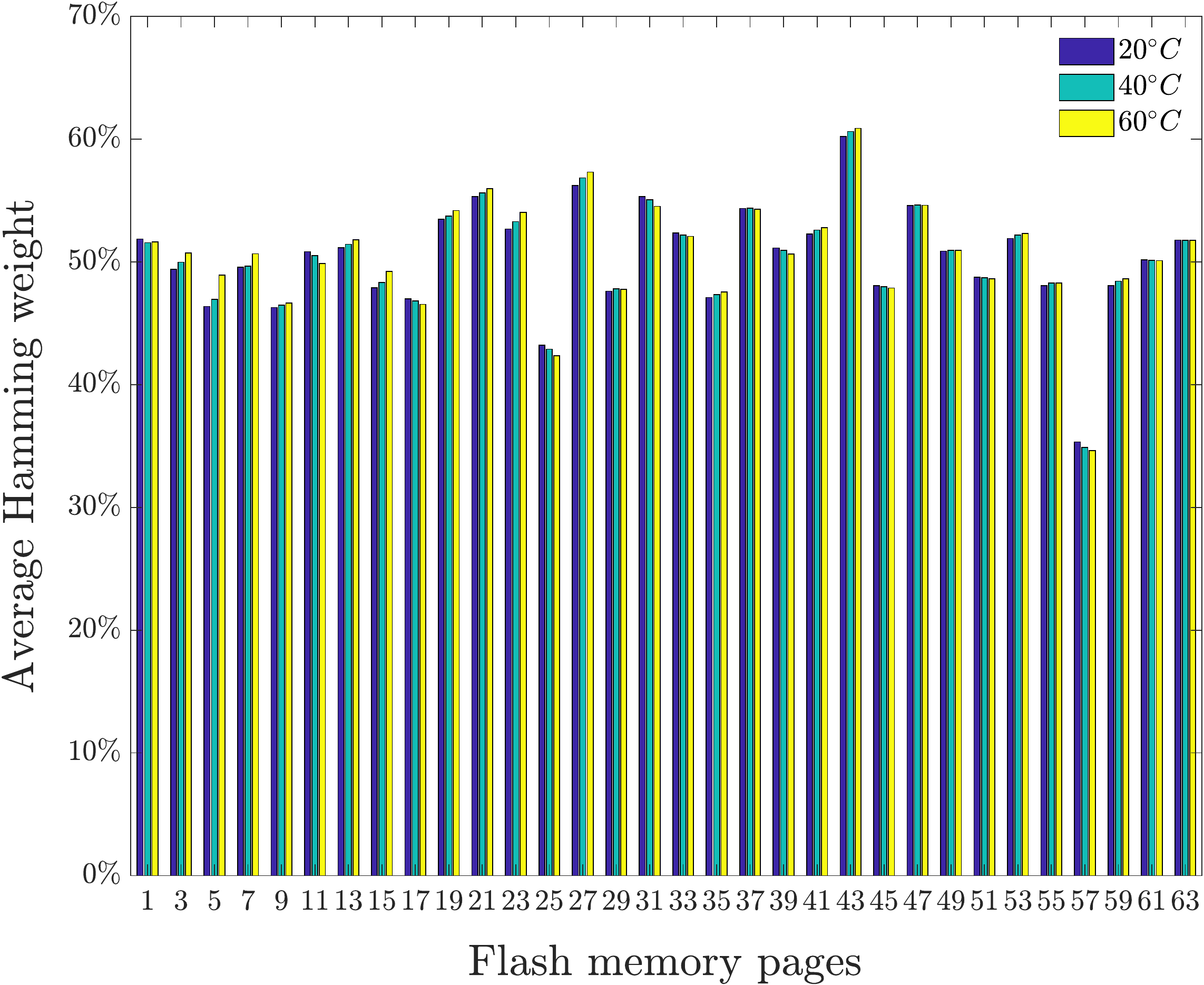}
\vspace{-2.5pt}
\caption{Average fractional Hamming weight values per instance of the examined NAND-Flash-memory-based {PUF}, for the different values of the ambient temperature tested.}
\label{fig:flashtempHWperpage}
\vspace{-5pt}
\end{figure}

As each memory block of the Samsung K9F1G08U0E NAND Flash memory is made up of 64 pages, 32 of these pages will constitute individual {PUF} instances, with 31 of them (pages `1', `3', `5', $\dotsc$, `61') receiving disturbances from both of their adjacent pages (double-side program ``hammering''), and one of them (page~`63') receiving disturbances from its only adjacent page (single-sided program ``hammering'', from page~`62'). For each page constituting a {PUF} instance, 20 responses have been received of a size of 2 {KB} each, for a total size of 64 {KB} per block measurement.

\begin{figure}[tb!]
\centering
\includegraphics[width=0.9\linewidth]{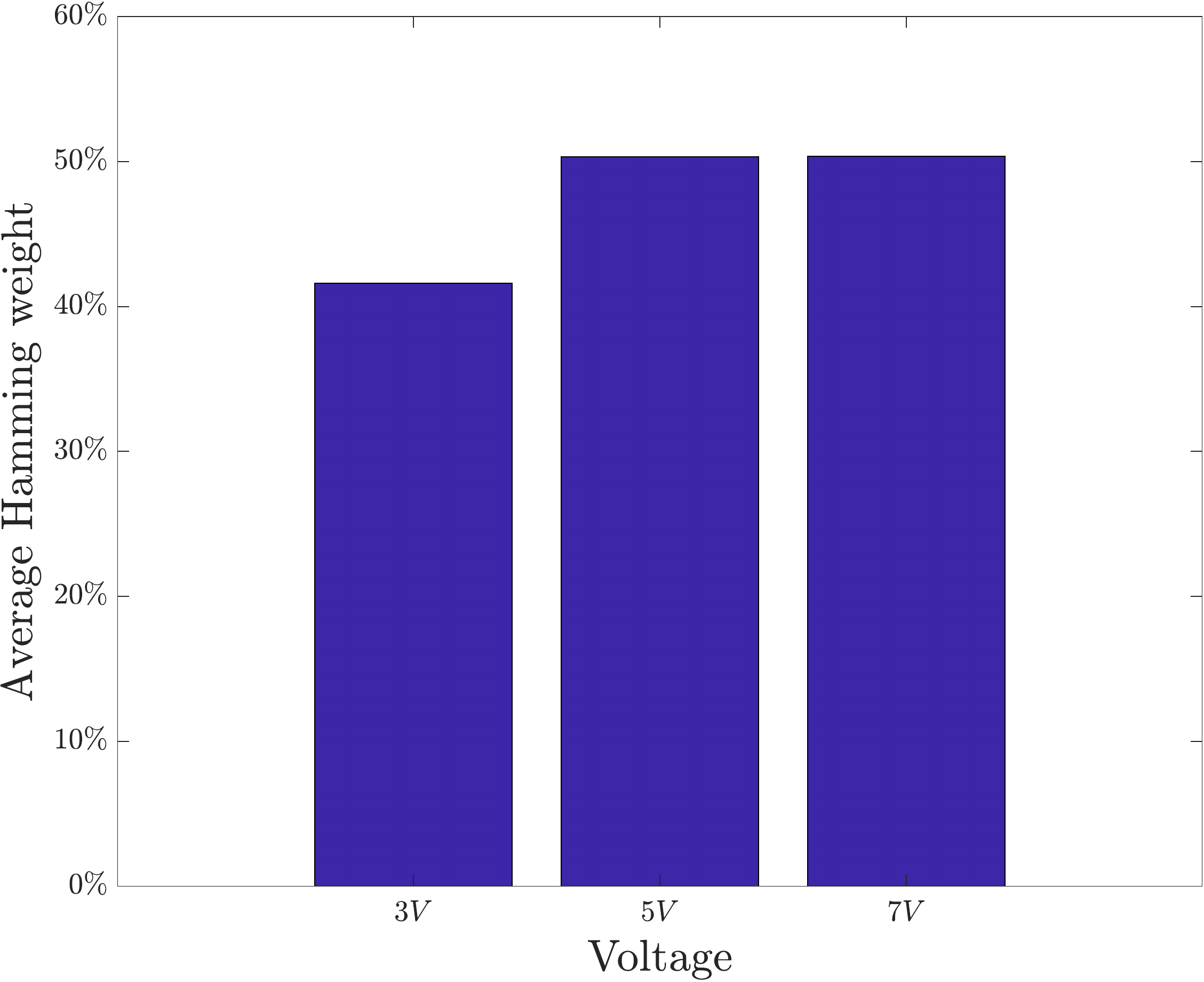}
\vspace{-2.5pt}
\caption{Overall average fractional Hamming weight value of the examined NAND-Flash-memory-based {PUF}, for each value of supply voltage provided by the {USB} port of the STM32F429I-DISC1 board to the overall system.}
\label{fig:flashusbHWoverall}
\vspace{-5pt}
\end{figure}

\begin{figure}[t!]
\centering
\includegraphics[width=0.9\linewidth]{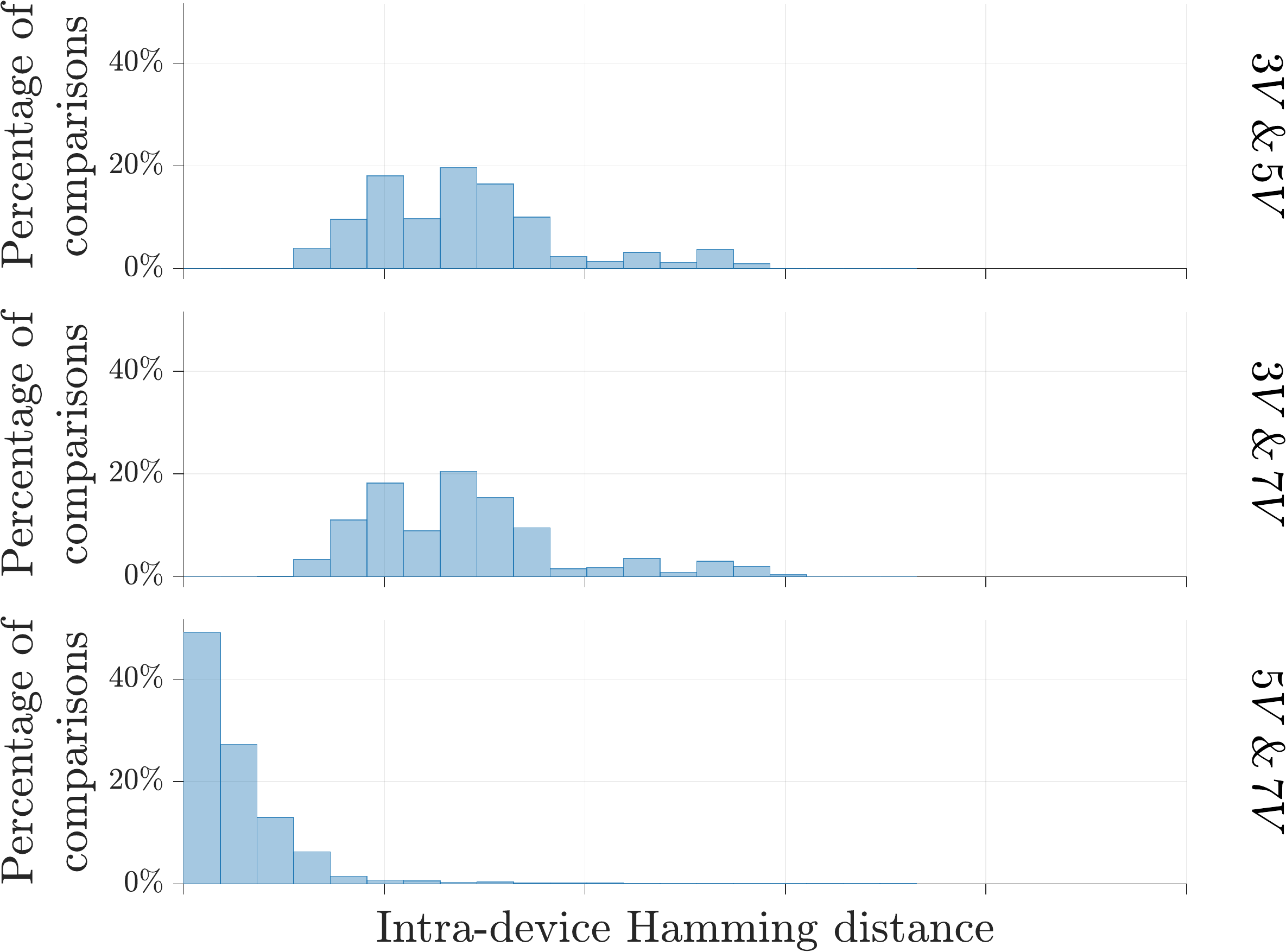}
\vspace{-2.5pt}
\caption{Intra-device Hamming distance values for the examined NAND-Flash-memory-based {PUF}, for pairs of {PUF} responses taken at different supply voltage levels provided by the {USB} port of the STM32F429I-DISC1 board to the overall system.}
\label{fig:flashusbinter}
\vspace{-10pt}
\end{figure}

Our results for measurements taken at $20${\degree}C, $40${\degree}C, and $60${\degree}C indicate an overall fractional Hamming weight of 50\% for the memory pages utilised as PUFs~(\Cref{fig:flashtempHWoverall}), but quite varying Hamming weights for each individual page serving as a PUF~(\Cref{fig:flashtempHWperpage}), clearly suggesting that it would be preferable to consider a whole memory block as a PUF, and not each individual page of such a block, as certain pages acting as PUFs appear to be biased towards one logical value or the other. Additionally, measurements taken at power supply voltages lower than the nominal lead to an overall Hamming weight value that is significantly below 50\%, indicating a bias towards the logical value of `0' and leading into intra-device Hamming distances significantly higher than 10\%, which may affect the ability of this PUF to act as a sustainable security mechanism. In this case, the employment of an internal voltage regulator, of a rather simple design, may potentially mitigate the observed effects of power supply voltage variations.

Therefore, although a number of issues that may reduce the practical applicability and sustainability of the examined NAND-Flash-memory-based PUFs have been identified, it may be possible to address them through rather realistic design modifications and improvements, which, however, remain to be implemented and tested in practice as part of future works on this scientific field. At the same time, however, we can also conclude that the examined PUFs are able to provide flexible and scalable security applications in a practical, liightweight, and highly sustainable manner, under nominal environmental conditions.

\bibliographystyle{./IEEEtran}
\balance
\bibliography{./IEEEabrv,./IEEEexample}

\end{document}